\documentclass[conference,final,a4paper]{IEEEtran}
\IEEEoverridecommandlockouts
\usepackage[utf8]{inputenc} 
\usepackage[T1]{fontenc}
\usepackage{ifthen}
\usepackage{cite}
\usepackage{times}

\usepackage{amsmath}  % Define \boldsymbol (in amsbsy too) and align
\usepackage{amssymb}  % Define \mathbb (in amsfonts too)

\usepackage{theorem}  % Helps in rendering theorems, etc.
\usepackage{cite}     % Gives multiple references as intervals
\usepackage{comment}  % Comments out with \begin{comment} ... \end{comment}

\usepackage{upref}
\usepackage{amsfonts}

\usepackage[dvipsnames,usenames]{color}
\usepackage{enumerate}

\usepackage{graphicx}

\usepackage{latexsym}

\usepackage{color}

\usepackage{multicol}
\usepackage{hhline}

\usepackage{psfrag}

\usepackage{tikz}
\usetikzlibrary{positioning, patterns, decorations.pathreplacing, calc}
\usepackage{pgfplots}

\usepackage{hhline}

\usepackage{algorithm}
\usepackage[noend]{algpseudocode}

\usepackage{xcolor,colortbl}

\usepackage{subcaption} 
\newcommand{\be}[1]{\begin{equation}\label{#1}}
\newcommand{\ee}{\end{equation}}

\newcommand{\bc}{\begin{center}}
\newcommand{\ec}{\end{center}}

\newcommand{\cC}{{\cal C}}

\newcommand{\cF}{{\cal F}}

\newcommand{\cL}{{\cal L}}

\newcommand{\bff}{{\boldsymbol f}}

\renewcommand{\geq}{\geqslant}

\newcommand{\Cref}[1]{Co\-rol\-la\-ry\,\ref{#1}}

\theoremstyle{plain} \theorembodyfont{\normalfont\slshape}

\newtheorem{thm}{Theorem$\!$}

\newtheorem{prop}[thm]{Proposition$\!$}

\newtheorem{lem}[thm]{Lemma$\!$}

\newtheorem{cor}[thm]{Corollary$\!$}

\newtheorem{defi}[thm]{Definition$\!$}

\theorembodyfont{\normalfont}

\newtheorem{remrk}{Remark$\!$}

\definecolor{Codecolor}{named}{White}  %{Tan}
\newcommand{\Copen}{\mbox{\{\kern-5.50pt\{}}
\newcommand{\Cclose}{\mbox{\}\kern-5.50pt\}}}
\newcommand{\Cslash}{\mbox{$\backslash\kern-6.02pt\backslash$}}

\newtheorem{innercustomgeneric}{\customgenericname}
\providecommand{\customgenericname}{}
\newcommand{\newcustomtheorem}[2]{%
\newenvironment{#1}[1]
{%
\renewcommand\customgenericname{#2}%
\renewcommand\theinnercustomgeneric{##1}%
\innercustomgeneric
}
{\endinnercustomgeneric}
}
\newcustomtheorem{customproposition}{Proposition}
\newcustomtheorem{customtheorem}{Theorem}
\newcustomtheorem{customlemma}{Lemma}
\newcustomtheorem{customclaim}{Claim}
\newcustomtheorem{customcorollary}{Corollary}
\usepackage{pdfpages}

\usepackage[bookmarks=false]{hyperref}

\usepackage[a4paper, left=1.6cm,  right=1.6cm, top=1.95cm, bottom=4.4cm]{geometry}

\usepackage{fancyhdr}

\begin{document}

\def\AllDefinitions{0}

\title{Coding Schemes for the Noisy Torn Paper Channel}\vspace{0ex}

\date{\today}
\author{
\IEEEauthorblockN{
    \textbf{Frederik Walter}\textsuperscript{*}, \textbf{Maria Abu-Sini}\textsuperscript{*}, \textbf{Nils Weinhardt}, \textbf{Antonia Wachter-Zeh}
}
\IEEEauthorblockA{
    School of Computation, Information and Technology, Technical University of Munich, DE-80333 Munich, Germany \\
    Emails: \{ {\it frederik.walter, maria.abu-sini, nils.weinhardt, antonia.wachter-zeh}\}@tum.de 
    \\[-5ex]}
	\thanks{* These authors contributed equally. Funded by the European Union (DiDAX, 101115134). Views and opinions expressed are however those of the author(s) only and do not necessarily reflect those of the European Union or the European Research Council Executive Agency. Neither the European Union nor the granting authority can be held responsible for them.}
}
\maketitle

\thispagestyle{fancy}

\begin{abstract}
    To make DNA a suitable medium for archival data storage, it is essential to consider the decay process of the strands observed in DNA storage systems. 
    This paper studies the decay process as a probabilistic noisy torn paper channel (TPC), which first corrupts the bits of the transmitted sequence in a probabilistic manner by substitutions, then breaks the sequence into a set of noisy unordered substrings.
    The present work devises coding schemes for the noisy TPC by embedding markers in the transmitted sequence. 
    We investigate the use of static markers and markers connected to the data in the form of hash functions. 
    These two tools have also been recently exploited to tackle the noiseless TPC.
    Simulations show that static markers excel at higher substitution probabilities, while data-dependent markers are superior at lower noise levels. Both approaches achieve reconstruction rates exceeding $99\%$ with no false decodings observed, primarily limited by computational resources.
\end{abstract}

\section{Introduction}
DNA strands used for data storage tend to decay over time, causing strands to break into short fragments~\cite{GrassChallenges}.
Moreover, DNA synthesis and sequencing processes are error-prone, introducing insertions, deletions, and substitutions.
Consequently, enabling long-term storage in such systems necessitates algorithms capable of recovering data from a set of disordered, noisy substrings.

The probabilistic binary noisy torn paper channel (TPC) models this decay and noise accumulation.
This channel takes a binary sequence, corrupts its bits uniformly at random with a substitution probability $p_s$, and subsequently breaks the noisy sequence at any position with probability $p_b = \frac{\alpha}{\log_2 n}$, where $n$ denotes the length of the transmitted sequence.
The capacity of this channel has been investigated in~\cite{Lit8}.
This paper devises two coding schemes for the noisy TPC, focusing on breakage and substitutions; the investigation of insertions and deletions is left for future work.
The two devised schemes build upon existing work~\cite{NestedHashing, ShomoronyFirstTornPaper}. 

A standard approach to reconstructing a sequence from unordered fragments involves interleaving a codeword of an erasure code with a marker followed by an index.
This technique is inspired by the \emph{interleaving pilot scheme} proposed in~\cite{ShomoronyFirstTornPaper}. 
However, a significant drawback of this scheme, as illustrated in Fig.~\ref{fig:ManyPermutations}, is that the interleaved indices are independent of the data. 
Breakages occurring between indices yield fragments devoid of position information, resulting in multiple fragment permutations that appear equally valid. 
To mitigate this ambiguity, \cite{NestedHashing,NestedVT} proposed interleaving the codeword with sequence-dependent information rather than fixed markers. 
Specifically, \cite{NestedHashing} suggested interleaving the sequence with hash values derived from data blocks.
Thus, the literature primarily employs two tools: (1) interleaving with markers and indices, and (2) interleaving with data-dependent hash values.

\subsection{Our Contribution}
In this work, we propose two coding schemes that leverage these tools to address the noisy TPC.
The first scheme, detailed in Section~\ref{sec:marker}, relies primarily on explicit indexing. It interleaves the sequence with markers (specifically $001$), followed by bits from a de Bruijn sequence acting as indices, and parity bits derived from the preceding data block. These parity bits serve to link the data with the structural redundancy.

The second scheme (Section~\ref{sec:nested_hash}) adopts a hash-based approach without explicit indices.
It interleaves the sequence with hash values designed to be robust against substitutions.
While the hash functions, as used in~\cite{NestedHashing}, are effective for the noiseless TPC, they are unsuitable here; a single substitution can drastically alter the hash value (the avalanche effect), making it impossible to verify the adjacency of two fragments.
To overcome this, we employ locality-sensitive hashing (LSH). LSH ensures that small changes in the data, such as those caused by substitutions, result in minimal changes to the hash value, thereby facilitating the correct identification of consecutive fragments even in the presence of noise.

\begin{figure}[htb]
    \centering
    \includegraphics[scale=0.25]{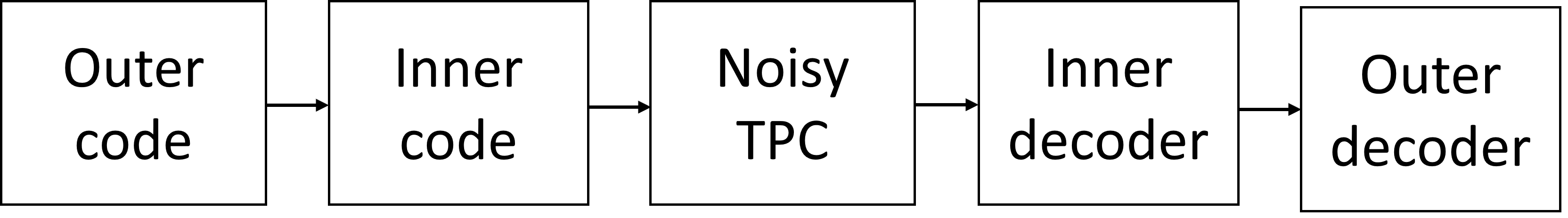}
    \caption{Concatenated coding scheme for the noisy TPC.}
    \label{Fig:ConcatenatedCode}
\end{figure}

To correct the substitutions after reconstructing the sequence, both schemes utilize an outer low-density parity-check (LDPC) code. The overall processing pipeline can be seen in Fig.~\ref{Fig:ConcatenatedCode}.

\begin{figure*}[t]
    \centering
    \includegraphics[scale=0.35]{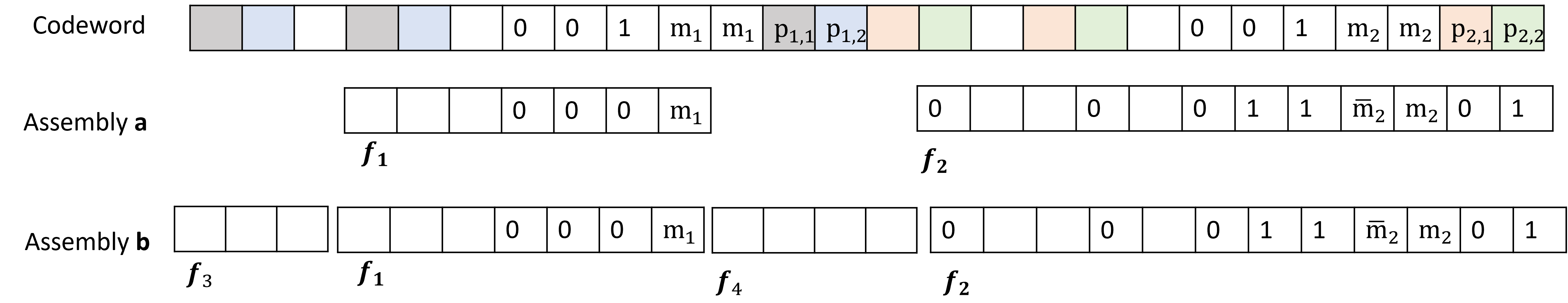}
    \caption{
    The structure of the coding scheme presented in Section~\ref{sec:marker} is illustrated in the upper word. Markers, followed by indices and parity bits are interleaved into an LDPC codeword. 
    The colors and the values inside the word and the fragments are referred to in Section~\ref{sec:marker}.
    Moreover, affixing the fragments $\bff_1$ and $\bff_2$ at indices $4$ and $15$, respectively, constitutes assembly $\mathbf{a}$, which is a partial assembly. Affixing fragments $\bff_3$ and $\bff_4$ in the gaps results in the complete assembly $\mathbf{b}$. \\[-6ex]}
    \label{fig:violationsCount}
\end{figure*}

Finally, Section~\ref{sec:simulation} presents simulations evaluating these schemes. We compare the robustness of static markers against data-dependent hashing to maintain reconstruction integrity and identify the optimal operating regimes for each approach.

\subsection{Further Related Work}
Beyond the discussed approaches, we highlight that~\cite{NestedVT} and~\cite{EmbeddedVT} introduced a coding scheme for the noiseless TPC utilizing Varshamov-Tenengolts (VT) codes as hash functions.
Furthermore, Ravi et al. studied in~\cite{ShomoronyEveryDistribution} and~\cite{Lit8} the capacity of the torn paper channel when fragments might be lost with a certain probability that depends on their lengths.
In parallel, adversarial settings of the TPC have been extensively studied~\cite{ISIT_2025, AdversarialTornPaper, Raviv2024_Arxiv, Raviv2024, Raviv2024_Practical, Lit10, GeneralizedTornPaper, Lit9, RavivSingleFrag_ISIT, RavivSingleFrag_Arxiv}. These works typically impose constraints such as minimum fragment lengths or upper bounds on the number of breakages and errors.
Finally, related models including shotgun sequencing and chop-and-shuffle channels have received significant attention in the recent literature as covered in~\cite{Lit1,Lit20, Lit2,Lit12,Lit3,Lit4,Lit5,Lit6,Lit7,Lit15,Lit14, Lit21, Lit22}. 
Details about the related channels are omitted due to lack of space.

\subsection{Notations}
Throughout this paper, we use $1$-based indexing. For a sequence $\mathbf{x}$, the notation $\mathbf{x}_{[i,j]}$ represents the substring starting at index $i$ and ending at index $j$ and $|\mathbf{x}|$ denotes the length. We denote by $\text{x}_j^i$ $i$ repetitions of the $j$-th bit of sequence $\mathbf{x}$. The operator $ \mathbf{x}_1 || \mathbf{x}_2$ denotes the concatenation of sequences $\mathbf{x}_1, \mathbf{x}_2$. Finally, an \textit{assembly} refers to the method of affixing fragments, at specific positions, as illustrated in Fig.~\ref{fig:violationsCount}.
We denote an assembly by small bold letters, and we call an assembly \emph{complete} if it covers the whole codeword. Otherwise, it is referred to as a \emph{partial assembly}.

\begin{figure}[htb]
    \centering
    \includegraphics[scale=0.3]{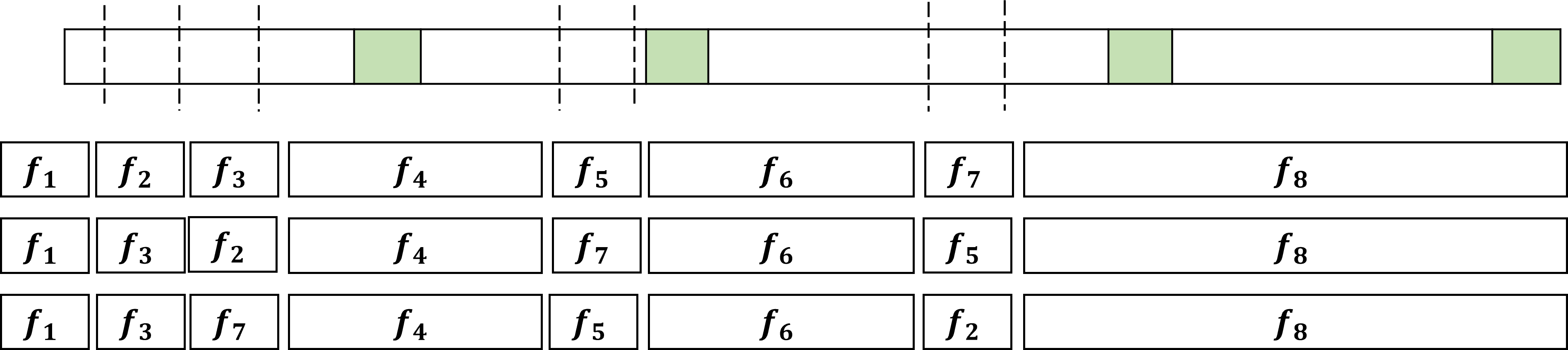}
    \caption{Markers (that are independent of the data) are highlighted above in green. Breakages between markers resulted in fragments $\bff_2, \bff_3, \bff_5$, and $\bff_7$. As illustrated, all possible permutations of these fragments result in a word that satisfies the markers and the indices constraints. Thus, they are all likely to be the correct permutation of the fragments.}
    \label{fig:ManyPermutations}
\end{figure}

\section{Marker-Based Codes with Indices}\label{sec:marker}

To tackle the noisy torn paper channel, we propose in this section a concatenated coding scheme as depicted in Fig.~\ref{Fig:ConcatenatedCode}. Subsection~\ref{Subsec:EncIndices} first presents the structure of the codewords, then Subsection~\ref{Subsec:DecIndices} elaborates on the reconstruction algorithm. 
The scheme has several parameters that control its redundancy, failure rate, and decoding run-time complexity. These parameters are $d, d', c_1, c_2$, and $k$, and their role will be explained in the remainder of the section.

\subsection{Encoding algorithm} \label{Subsec:EncIndices}

The outer code used in this scheme is an LDPC code that can correct substitutions. The inner code then interleaves markers and indices in an LDPC codeword. Namely, let $\mathbf{x}$ be a message we aim to transmit, and denote by $\mathbf{x}'$ an LDPC codeword obtained from encoding $\mathbf{x}$. Furthermore, let $\mathbf{r}$ be a fixed word chosen uniformly at random. We denote by $\mathbf{x}'' \triangleq \mathbf{x}'+ \mathbf{r}$, where the summation is taken over the binary field. 
The importance of this summation will be explained at the end of the section.

The overall codeword is denoted as $ f\left(\mathbf{x} \right) $, and is constructed from the concatenation of $\frac{\left|\mathbf{x}'' \right|}{d}$ sequences, which we will refer to as \emph{encoded blocks}. That is,
\begin{equation}\label{eq:codeword}
f \left( \mathbf{x} \right)  \triangleq
\mathbf{b}_1 \parallel\mathbf{b}_2 \parallel\cdots \parallel \mathbf{b}_{\frac{\left|\mathbf{x}'' \right|}{d}},
\end{equation}
and the blocks $\mathbf{b}_i$'s are defined by
\begin{equation}\label{eq:block}
\mathbf{b}_i \triangleq \mathbf{x}''_{[(i-1)d+1,id]} \parallel 001 {\text{m}}_i^{c_1} \text{p}_{i,1} \cdots \text{p}_{i,c_2},
\end{equation}
where
\begin{itemize}
    \item $\mathbf{m} = {\text{m}}_1 \cdots {\text{m}}_{\frac{\left| \mathbf{x}'' \right|}{d}}$ is a prefix of a shortest de Bruijn sequence that is still longer than $\frac{\left|\mathbf{x}'' \right|}{d}$,
    \item $\text{m}_i^{c_1}$ denotes $c_1$ repetitions of the $i$-th bit of $\mathbf{m}$, and
    \item ${\text{p}}_{i,j}$ is the parity of $\frac{d}{d'}$ bits in $\mathbf{x}''_{\left[ (i-1)d+1, id \right]}$, specifically of the bits $ \text{x}''_{(i-1)d+j}, \text{x}''_{(i-1)d+j+d'}, \ldots, \text{x}''_{d-d'+j} $, which are located at regular intervals of distance $d'$.
\end{itemize}
For simplicity we choose parameters $d$ and $d'$ such that $d$ divides $\left| \mathbf{x}'' \right|$ and $d'$ divides $d$.
In summary, each encoded block $\mathbf{b}_i$ is obtained from concatenating the length-$d$ substring $\mathbf{x}''_{[(i-1)d+1, id]}$, a marker $001$, $c_1$ repetitions of the $i$-th bit of a de Bruijn sequence $\mathbf{m}$, and lastly parity bits on the substring $\mathbf{x}''_{[(i-1)d+1,id]}$.
As an example,
the structure of the overall codeword is illustrated in Fig.~\ref{fig:violationsCount} when the scheme's parameters are $\left| \mathbf{x}''\right| = 12, c_1=2, c_2 = 2$, and $d' = 3$.
Specifically, a parity bit ${\text{p}}_{i,j}$ and its corresponding bits in $\mathbf{x}''$ are highlighted in the same color in Fig.~\ref{fig:violationsCount}. We also emphasize that the choice of the marker $001$, over $000$ for instance, is based on the  periodicity property as discussed in~\cite{Sellers1962, Weindl2007}. 

As later demonstrated in Subsection~\ref{Subsec:DecIndices}, the key idea behind this coding scheme is using the de Bruijn bits as indices to determine the locations of long fragments. To enable correct detection of these indices in a fragment (even in the presence of substitution errors) we 
do not insert $\text{m}_i$ solely, but a repetition $\text{m}^{c_1}_i$ preceded by the marker $001$.
This protects the indices from substitutions, and hence extends the interleaving pilot scheme proposed by Shomorony et al.~\cite{ShomoronyFirstTornPaper} to the noisy TPC as well.
The parity bits $\text{p}_{i,j}$'s connect the redundancy with the preceding substring ${\mathbf{x}}''_{[(i-1)d, id]}$, and hence resolve the challenge exhibited in Fig.~\ref{fig:ManyPermutations}. Another approach to connect the redundancy with the data is to use hash functions, as proposed in Section~\ref{sec:nested_hash}.
We notice here that larger $c_2$ and smaller $d'$ (i.e., more hash bits over smaller number of data bits) result in stronger hash functions which can handle more breakages and errors. However, larger $c_2$ increases the scheme's redundancy as well. In general, in our simulations, we always try to find the smaller value of $c_2$ that still ensures acceptable failure rate.

Lastly, the significance of adding the random word $\mathbf{r}$ is to reduce the occurrences of the marker $001$ in $\mathbf{x''}$, and hence enable correct detection of the markers in the noisy fragments.

\subsection{Decoding Algorithm} \label{Subsec:DecIndices}

To explain the decoding algorithm, notice first that the markers, indices, and parities behave like constraints on the codeword. The simplest possible decoding algorithm is one that iterates over all possible permutations of the fragments, and checks how many constraints each permutation violates. 
A constraint is violated if, for instance, we expect a marker $001$ but have another substring such as $011$, or when a bit ${\text{p}}_{i,j}$ does not equal the parity of the corresponding bits in $\mathbf{x}_{[(i-1)d+1, id]}$.
The permutation with the fewest violated constraints is most likely to be the correct one. Therefore, we define the \emph{violations count} of an assembly to be the number of constraints (imposed by markers, indices, and parities) it violates.
For instance, the violations count of assembly $\mathbf{a}$ depicted in Fig.~\ref{fig:violationsCount} equals $4$ due to the following violations.
\begin{itemize}
    \item Since marker $001$ is used, the $9$-th and the $21$-st bits are expected to be $1$ and $0$, respectively.
    \item The $23$-rd bit should be assigned the value $m_2$ according to the used de-Bruijn sequence.
    \item As illustrated in the figure, $\text{p}_{2,2}$ is the parity of the $15$-th and the $18$-th bits, in which we have $0$'s, thus we expect to have $0$ in the $26$-th bit.
\end{itemize}

Due to the infeasible run-time complexity of iterating over all permutations, we propose a beam search reconstruction algorithm which operates in the following three phases. Beforehand, denote the input to the decoder, i.e., the received set of $t$ noisy fragments, by $\cF = \left\{ \bff_1, \bff_2, \ldots, \bff_t \right\}$.

\emph{Phase 1:}
Define $ S_1 \triangleq \left\{ \bff_i : \left|\bff_i \right| \geq\ell \cdot d_{enc}\right\} $ to be a set of sufficiently long fragments. The parameter $\ell$ determines here which fragment's length is considered to be long enough.
Since the fragments of $S_1$ are sufficiently long, they contain multiple markers, and hence a sufficiently long substring of $\mathbf{m}$ that enables detecting their locations. Therefore in Phase $1$ we find all possible locations of each fragment $\bff_i \in S_1$, and accordingly generate assemblies that affix all of the fragments of $S_1$. 
These assemblies are then ranked according to their violations count in an ascending order and the first $k$ are stored in heap $A$ as the initial beams.

\emph{Phase 2:}
In this phase we extend the assemblies in $A$ into complete ones. 
To that end, we consider each assembly $\mathbf{a} \in A$ and try to extend it by affixing some fragment adjacent to the first attached one in $\mathbf{a}$. This results in many possible extensions of the assemblies in $A$. We choose the best $k$ extensions and store them according to their violations count in an ascending order in $A$. We keep repeating this phase until a sufficient number of complete assemblies is found or until $A$ becomes empty. $A$ becomes empty if at some point we could not extend any partial assembly in it, namely we stored wrong partial assemblies in the previous iterations of Phase $2$.
Moreover, the parameter $k$ determines the size of the heap $A$. The larger it is, the more partial assemblies are stored at a time, and hence the higher the run-time complexity of Phase $2$ is, but the smaller the failure rate of the decoding algorithm gets.

\emph{Phase 3:}
In Phase $2$ we collect a certain number of complete assemblies. Next, in Phase $3$ we extract the corrupted bits from the LDPC codeword $\mathbf{x}''$ and apply $100$ iterations of the min-sum decoding algorithm on these complete assemblies. The output of this phase is a set $\cL$. 
Whenever the $100$ iterations of the min-sum decoder converge to an LDPC codeword, we add it to the set $\cL$. The reconstruction is considered to be successful if at the end $\cL$ consists of the single correct message we aimed to transmit.
In the simulations we observed that all failures are due to an empty set $\cL$ rather than $\cL$ consisting of multiple words. That is, even when we applied the min-sum decoder on a wrong assembly, the $100$ iterations did not converge to a wrong LDPC codeword. The failure reason we mostly focused on in this work is when $A$ is empty or does not contain the correct assembly.

\section{Nested Codes without Indices}\label{sec:nested_hash}

For moderate codeword lengths with few fragments, explicit indices are unnecessary. Instead, we locate the longest fragment using markers and fragment lengths, then employ beam search to concatenate the rest.

\subsection{Encoding Scheme}

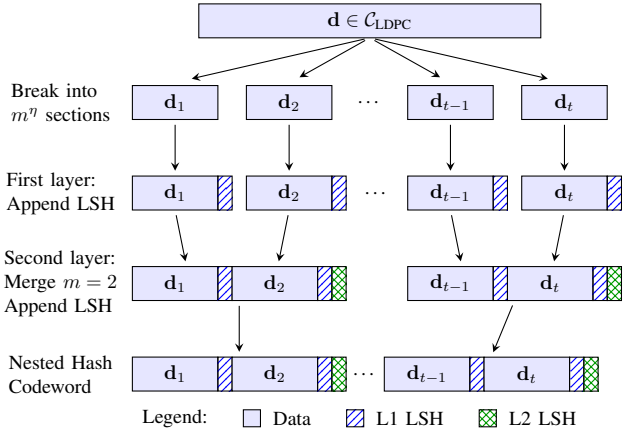
\begin{figure}
    \centering
    \begin{tikzpicture}[scale=0.65, transform shape,
    box/.style={
        draw, 
        rectangle, 
        minimum height=0.6cm, 
        fill=blue!10, 
        outer sep=0pt,
        inner sep=0pt
        },
        hash1/.style={
        draw, 
        rectangle, 
        minimum height=0.6cm, 
        pattern=north east lines, 
        pattern color=blue, 
        outer sep=0pt
        },
        hash2/.style={
        draw, 
        rectangle, 
        minimum height=0.6cm, 
        pattern=crosshatch, 
        pattern color=green!60!black, 
        outer sep=0pt
        },
        hash3/.style={
        draw, 
        rectangle, 
        minimum height=0.6cm, 
        pattern=north west lines, 
        pattern color=red, 
        outer sep=0pt
        },
        arrow/.style={->, >=stealth, shorten >=2pt, shorten <=2pt}
]

    % --- Level 0: Data / Outer LDPC ---
        \node[box, minimum width=6cm] (data) {$\mathbf{d} \in \cC_{\text{LDPC}}$};

        % --- Level 1: Break into Sections ---
        \coordinate (split1) at (data.south west);
        \coordinate (split2) at (data.south east);
        
        % Define widths
        \def\secwidth{1.5cm}
        \def\hashwidth{0.25cm}
        \def\textboxwidth{2.5cm}
        \def\hdist{0.7cm}
        
        % Draw 4 sections representing data blocks
        \node[below=\hdist of data.south] (dots1) {\ldots};
        \node[box, minimum width=\secwidth, left=1cm of dots1.east] (sec2) {$\mathbf{d}_2$};
        \node[box, minimum width=\secwidth, left=0.5cm of sec2] (sec1) {$\mathbf{d}_1$};
        \node[box, minimum width=\secwidth, right=1cm of dots1.west] (sec3) {$\mathbf{d}_{t-1}$};
        \node[box, minimum width=\secwidth, right=0.5cm of sec3] (sec4) {$\mathbf{d}_{t}$};
        
        \node[left=0cm of sec1, align=left, minimum width=\textboxwidth] {Break into \\ $m^{\eta}$ sections};
        
        % Arrows from Data to Sections
        \draw[arrow] (data.south) -- ($(sec1.north) + (0.2,0.1)$);
        \draw[arrow] (data.south) -- ($(sec2.north) + (0.1,0.1)$);
        \draw[arrow] (data.south) -- ($(sec3.north) + (-0.1,0.1)$);
        \draw[arrow] (data.south) -- ($(sec4.north) + (-0.2,0.1)$);
        
        % --- Level 2: First Layer (LSH) ---
        % Section 1 + LSH
        \node[box, minimum width=\secwidth, below=\hdist of sec1] (l1_sec1) {$\mathbf{d}_1$};
        \node[hash1, minimum width=\hashwidth, right=0cm of l1_sec1, label=center:{}] (l1_h1) {};
        
        % Section 2 + LSH
        \node[box, minimum width=\secwidth, below=\hdist of sec2] (l1_sec2) {$\mathbf{d}_2$};
        \node[hash1, minimum width=\hashwidth, right=0cm of l1_sec2] (l1_h2) {};
        
        % Section 3 + LSH
        \node[box, minimum width=\secwidth, below=\hdist of sec3] (l1_sec3) {$\mathbf{d}_{t-1}$};
        \node[hash1, minimum width=\hashwidth, right=0cm of l1_sec3] (l1_h3) {};
        
        % Section 4 + LSH
        \node[box, minimum width=\secwidth, below=\hdist of sec4] (l1_sec4) {$\mathbf{d}_{t}$};
        \node[hash1, minimum width=\hashwidth, right=0cm of l1_sec4] (l1_h4) {};

        % Dots
        \node at ($(l1_h2.east)!0.5!(l1_sec3.west)$) {\ldots};

        \node[left=0cm of l1_sec1, align=left, minimum width=\textboxwidth] {First layer:\\ Append LSH};
        
        % Arrows
        \draw[arrow] (sec1) -- (l1_sec1);
        \draw[arrow] (sec2) -- (l1_sec2);
        \draw[arrow] (sec3) -- (l1_sec3);
        \draw[arrow] (sec4) -- (l1_sec4);

        % --- Level 3: Second Layer (Merge + Hash) ---
        % Merge 1 & 2
        \node[box, minimum width=\secwidth, below=\hdist of l1_sec1, xshift=0cm] (l2_p1) {$\mathbf{d}_1$};
        \node[hash1, minimum width=\hashwidth, right=0cm of l2_p1] (l2_h1_inner) {};
        \node[box, minimum width=\secwidth, right=0cm of l2_h1_inner] (l2_p2) {$\mathbf{d}_2$};
        \node[hash1, minimum width=\hashwidth, right=0cm of l2_p2] (l2_h2_inner) {};
        \node[hash2, minimum width=\hashwidth, right=0cm of l2_h2_inner] (l2_Hash2_1) {};
        
        % Merge 3 & 4
        \node[box, minimum width=\secwidth, below=\hdist of l1_sec3, xshift=0cm] (l2_p3) {$\mathbf{d}_{t-1}$};
        \node[hash1, minimum width=\hashwidth, right=0cm of l2_p3] (l2_h3_inner) {};
        \node[box, minimum width=\secwidth, right=0cm of l2_h3_inner] (l2_p4) {$\mathbf{d}_{t}$};
        \node[hash1, minimum width=\hashwidth, right=0cm of l2_p4] (l2_h4_inner) {};
        \node[hash2, minimum width=\hashwidth, right=0cm of l2_h4_inner] (l2_Hash2_2) {};

        \node[left=0cm of l2_p1, align=left, minimum width=\textboxwidth] {Second layer:\\ Merge $m=2$ \\ Append LSH};

        % Arrows for merge
        \draw[arrow] (l1_sec1.south) -- ++(0.2,-\hdist);
        \draw[arrow] (l1_sec2.south) -- ++(-0.2,-\hdist);
        
        \draw[arrow] (l1_sec3.south) -- ++(0.2,-\hdist);
        \draw[arrow] (l1_sec4.south) -- ++(-0.2,-\hdist);
        % --- Level 4: Final Codeword (Third Layer) ---
        % Final Concatenation
        \node[box, minimum width=\secwidth, below=\hdist of l2_p1] (l3_p1) {$\mathbf{d}_1$};
        \node[hash1, minimum width=\hashwidth, right=0cm of l3_p1] (l3_h1) {};
        \node[box, minimum width=\secwidth, right=0cm of l3_h1] (l3_p2) {$\mathbf{d}_2$};
        \node[hash1, minimum width=\hashwidth, right=0cm of l3_p2] (l3_h2) {};
        \node[hash2, minimum width=\hashwidth, right=0cm of l3_h2] (l3_h3) {};

        % Dots
        \node[minimum width=0.5cm, right=0cm of l3_h3] (l3_dots) {\ldots};
        
        \node[box, minimum width=\secwidth, right=0cm of l3_dots] (l3_p3) {$\mathbf{d}_{t-1}$};
        \node[hash1, minimum width=\hashwidth, right=0cm of l3_p3] (l3_h4) {};
        \node[box, minimum width=\secwidth, right=0cm of l3_h4] (l3_p4) {$\mathbf{d}_{t}$};
        \node[hash1, minimum width=\hashwidth, right=0cm of l3_p4] (l3_h5) {};
        \node[hash2, minimum width=\hashwidth, right=0cm of l3_h5] (l3_h6) {};
        
        % \node[hash3, minimum width=\hashwidth, right=0cm of l3_h6] (l3_h7) {};

        \node[left=0cm of l3_p1, align=left, minimum width=\textboxwidth] {Nested Hash \\ Codeword};

        % Arrows to final
        \draw[arrow] ($(l2_p1.south west)!0.5!(l2_Hash2_1.south east)$) -- ($(l3_p1.north west)!0.5!(l3_h3.north east)$);
        \draw[arrow] ($(l2_p3.south west)!0.5!(l2_Hash2_2.south east)$) -- ($(l3_p3.north west)!0.5!(l3_h6.north east)$);

        % Legend (fixed)
        \node[anchor=west, below=0.2cm of l3_p1] (legend) {Legend:};
        \node[box, minimum size=0.3cm, right=0.5cm of legend] (legData) {};
        \node[right=0.1cm of legData] (data) {Data};
        \node[hash1, minimum size=0.3cm, right=0.5cm of data] (legL1) {};
        \node[right=0.1cm of legL1] (L1) {L1 LSH};
        \node[hash2, minimum size=0.3cm, right=0.5cm of L1] (legL2) {};
        \node[right=0.1cm of legL2] (L2) {L2 LSH};

\end{tikzpicture}
    \caption{Nested encoding with $L=3$ layers and branching factor $m=2$. Each block contains $d$ data bits and $p_\ell$ parity bits at layer $\ell$. \\[-6ex]}
    \label{fig:nested_hash}
\end{figure}

We adapt the nested reassembly codes of \cite{NestedHashing} for the noisy TPC. The encoder uses $L$ layers with branching factor $m$. At layer $\ell=0$, data is split into $m^{L-1}$ blocks of $d$ bits. A marker $\mathbf{p}_\ell \in \{0,1\}^{p_\ell}$ follows each block $i$:
    $\mathbf{c}_i^{(\ell)} = \mathbf{d}_i^{(\ell)} \,\|\, \mathbf{p}_\ell $ ,
where $\mathbf{d}_i^{(\ell)}$ is the data of block $i$ at layer $\ell$. Higher layers $\ell > 0$ concatenate $m$ layer-$(\ell-1)$ codewords to form data blocks:
\vspace{-3ex}
\begin{equation}
    \mathbf{d}_i^{(\ell)} = \mathbf{c}_{mi}^{(\ell-1)} \,\|\, \mathbf{c}_{mi+1}^{(\ell-1)} \,\|\, \cdots \,\|\, \mathbf{c}_{mi+m-1}^{(\ell-1)}.
\end{equation}
The final codeword $\mathbf{x}$ corresponds to layer $L-1$. No top-level parity is needed as the outer LDPC code validates the assembly (see Fig.~\ref{fig:nested_hash}).
The nested code rate is
\vspace{-0.2cm}
\begin{equation}
    R_{\text{Nested}} = \frac{m^{L-1} \cdot d}{\sum_{\ell=0}^{L-1} m^{L-1-\ell} \cdot (m^\ell \cdot d + p_\ell)},
\end{equation}
yielding a total rate $R = R_{\text{LDPC}} \cdot R_{\text{Nested}}$.

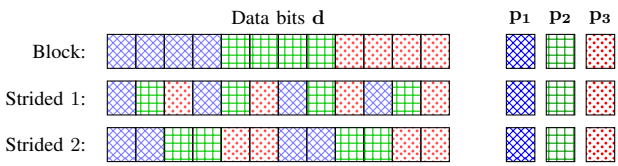
\begin{figure}
    \centering
    \begin{tikzpicture}[scale=0.65, transform shape,
    data1/.style={
        draw, 
        rectangle, 
        minimum height=0.6cm, 
        fill=blue!10, 
        pattern=crosshatch, 
        pattern color=blue!50,
        outer sep=0pt,
        inner sep=0pt
        },
    data2/.style={
        draw, 
        rectangle, 
        minimum height=0.6cm, 
        fill=green!10, 
        pattern=grid, 
        pattern color=green!80!black,
        outer sep=0pt,
        inner sep=0pt
        },
    data3/.style={
        draw, 
        rectangle, 
        minimum height=0.6cm, 
        fill=red!10, 
        pattern=crosshatch dots, 
        pattern color=red!70,
        outer sep=0pt,
        inner sep=0pt
        },
    hash1/.style={
        draw, 
        rectangle, 
        minimum height=0.6cm, 
        fill=blue!10, 
        pattern=crosshatch, 
        pattern color=blue,
        outer sep=0pt
        },
    hash2/.style={
        draw, 
        rectangle, 
        minimum height=0.6cm, 
        fill=green!10, 
        pattern=grid, 
        pattern color=green!60!black, 
        outer sep=0pt
        },
    hash3/.style={
        draw, 
        rectangle, 
        minimum height=0.6cm, 
        fill=red!10, 
        pattern=crosshatch dots, 
        pattern color=red,
        outer sep=0pt
        },
        arrow/.style={->, >=stealth, shorten >=2pt, shorten <=2pt}
]

    \def\bitwidth{0.5cm}
    % --- Contiguous ---
        \node[] (d1_0) {};
        \foreach \i in {1,2,3,4} {
            \node[data1, minimum width=\bitwidth, right=0cm of d1_\the\numexpr\i-1\relax] (d1_\i) {};
        }
        \foreach \i in {5,6,7,8} {
            \node[data2, minimum width=\bitwidth, right=0cm of d1_\the\numexpr\i-1\relax] (d1_\i) {};
        }
        \foreach \i in {9,10,11,12} {
            \node[data3, minimum width=\bitwidth, right=0cm of d1_\the\numexpr\i-1\relax] (d1_\i) {};
        }

        \node[hash1, minimum width=\bitwidth, right=1cm of d1_12] (h1_1) {};
        \node[hash2, minimum width=\bitwidth, right=0.2cm of h1_1] (h2_1) {};
        \node[hash3, minimum width=\bitwidth, right=0.2cm of h2_1] (h3_1) {};

        \node[align=right, left=0cm of d1_0.west] (d1_text) {Block:};

    % --- Legend above ---
        \node[] (data) at ($(d1_6.east) + (0,0.6cm)$) {Data bits $\mathbf{d}$};
        \node[] (red1) at ($(h1_1) + (0,0.6cm)$) {$\mathbf{p_1}$};
        \node[] (red2) at ($(h2_1) + (0,0.6cm)$) {$\mathbf{p_2}$};
        \node[] (red3) at ($(h3_1) + (0,0.6cm)$) {$\mathbf{p_3}$};

    % --- Strided p_l ---
        \node[] (d2_0) at ($(d1_0.south) + (0,-0.7cm)$) {};
        \foreach \i in {1,4,7,10} {
            \node[data1, minimum width=\bitwidth, right=0cm of d2_\the\numexpr\i-1\relax] (d2_\i) {};
            \node[data2, minimum width=\bitwidth, right=0cm of d2_\the\numexpr\i\relax] (d2_\the\numexpr\i+1\relax) {};
            \node[data3, minimum width=\bitwidth, right=0cm of d2_\the\numexpr\i+1\relax] (d2_\the\numexpr\i+2\relax) {};
        }  
        \node[hash1, minimum width=\bitwidth, right=1cm of d2_12] (h2_1) {};
        \node[hash2, minimum width=\bitwidth, right=0.2cm of h2_1] (h2_2) {};
        \node[hash3, minimum width=\bitwidth, right=0.2cm of h2_2] (h2_3) {};  
        
        \node[align=right, left=0cm of d2_0.west] (d2_text) {Strided $1$:};

    % --- Strided 2 p_l ---
        \node[] (d3_0) at ($(d2_0.south) + (0,-0.7cm)$) {};
        \foreach \i in {1,7} {
            \node[data1, minimum width=\bitwidth, right=0cm of d3_\the\numexpr\i-1\relax] (d3_\i) {};
            \node[data1, minimum width=\bitwidth, right=0cm of d3_\the\numexpr\i\relax] (d3_\the\numexpr\i+1\relax) {};
            \node[data2, minimum width=\bitwidth, right=0cm of d3_\the\numexpr\i+1\relax] (d3_\the\numexpr\i+2\relax) {};
            \node[data2, minimum width=\bitwidth, right=0cm of d3_\the\numexpr\i+2\relax] (d3_\the\numexpr\i+3\relax) {};
            \node[data3, minimum width=\bitwidth, right=0cm of d3_\the\numexpr\i+3\relax] (d3_\the\numexpr\i+4\relax) {};
            \node[data3, minimum width=\bitwidth, right=0cm of d3_\the\numexpr\i+4\relax] (d3_\the\numexpr\i+5\relax) {};
        }  
        \node[hash1, minimum width=\bitwidth, right=1cm of d3_12] (h3_1) {};
        \node[hash2, minimum width=\bitwidth, right=0.2cm of h3_1] (h3_2) {};
        \node[hash3, minimum width=\bitwidth, right=0.2cm of h3_2] (h3_3) {};  
        
        \node[align=right, left=0cm of d3_0.west] (d3_text) {Strided $2$:};

\end{tikzpicture}
    \caption{Subset selection schemes for Locality-Sensitive Hashing (LSH). Each parity bit on the right is derived from a majority vote over the similarly highlighted data bits on the left. For example the first parity bit in the Block scheme (blue) is the majority vote of the first $4$ data bits (also blue). \\[-6ex]}
    \label{fig:lsh}
\end{figure}

We replace static markers with Locality-Sensitive Hash (LSH) values, computed as majority votes over data bit subsets. Unlike the hashes with high avalanche effect used in \cite{NestedHashing}, this LSH is robust to small errors. 
We compare three subset selection schemes (Fig.~\ref{fig:lsh}):
\begin{itemize}
    \item \textbf{Block (Block):} Contiguous data blocks.
    \item \textbf{Stride 1 (LSH1):} Bits at regular intervals.
    \item \textbf{Stride 2 (LSH2):} Larger stride to reduce shift sensitivity.
\end{itemize}
An illustration of the different subset selection schemes for a data block of length $d=12$ is provided in Fig.~\ref{fig:lsh}. A comparison of their performance in the decoding process is presented in Section~\ref{sec:simulation}. 
The Block scheme struggles when fragments are appended at the beginning with approximately one block overlap, as that majority vote of a wrongly concatenated block has a $50\%$ chance of being correct anyway. In the Stride~1 scheme, the influence of a concatenated fragment is distributed among more parity bits. However, Stride~1 is sensitive to shifts within the codeword. Stride~2 offers a middle ground by using a larger stride to reduce shift sensitivity. Our simulations confirm a slight improvement in reassembly performance over Stride~1.

\subsection{Nested Hash Decoding}

We reconstruct $\mathbf{x}$ from fragments $\mathcal{F}$ using a two-phase beam search.

\emph{Phase 1: Beam Initialization.}
We start with the longest fragment $\bff_{\max} \in \mathcal{F}$. Its start position must correspond to the cumulative length of a subset of remaining fragments $\mathcal{L}$:
\begin{equation}
    \mathcal{S} = \left\{ s : s = \sum_{L \in \mathcal{I}} L, \, \mathcal{I} \subseteq \mathcal{L}  \right\}.
\end{equation}
For each $s \in \mathcal{S}$, we compute the parity distance $\Delta(\bff_{\max}, s)$ by comparing recomputed hashes to stored bits. Positions with equal distance are grouped into beams.

\emph{Phase 2: Beam Search Assembly.}
We maintain beams $\mathcal{B}$, each being a partial assembly with a parity distance. Iteratively, the best beam is expanded by prepending or appending unused fragments, and new distances are computed. We retain the best $B_{\max}$ beams.
Upon finding a complete assembly, we apply LDPC belief propagation. If successful, the codeword is returned; otherwise, the search continues.

\begin{algorithm}[t]
\caption{Beam Search Fragment Assembly}\label{alg:assembly}
\begin{algorithmic}[1]
\Require Fragment set $\mathcal{F}$, codeword length $n$, max beams $B_{\max}$
\Ensure Assembled codeword $\hat{\mathbf{x}}$ or failure
\State $f_{\max} \gets$ longest fragment in $\mathcal{F}$
\State $\mathcal{L} \gets$ multiset of lengths in $\mathcal{F} \setminus \{f_{\max}\}$
\State $\mathcal{S} \gets \{ s : s = \sum_{i \in I} \ell_i, \, I \subseteq \mathcal{L} \}$ \Comment{Subset sums}
\For{each $s \in \mathcal{S}$}
    \State Compute parity distance $\Delta(f_{\max}, s)$
\EndFor
\State Group positions by distance into beams $\mathcal{B} = \{B_1, B_2, \ldots\}$
\While{$\mathcal{B} \neq \emptyset$}
    \State Pop beam $B$ with lowest parity distance from $\mathcal{B}$
    \If{$B$ spans full length $n$}
        \State Decode with LDPC decoder
        \If{valid codeword found}
            \State \Return decoded codeword
        \EndIf
    \EndIf
    \For{each unused fragment $\bff_j \in \mathcal{F} \setminus B$}
        \For{$op \in \{\text{prepend}, \text{append}\}$}
            \State $B' \gets$ apply $op$ using $\bff_j$ on $B$
            \State Compute parity distance for all positions in $B'$
            \If{$|\mathcal{B}| < B_{\max}$ \textbf{or} $\Delta(B') < \max_{B_i \in \mathcal{B}} \Delta(B_i)$}
                \State Add $B'$ to $\mathcal{B}$ (remove worst beam if at capacity)
            \EndIf
        \EndFor
    \EndFor
\EndWhile
\State \Return failure
\end{algorithmic}
\end{algorithm}

\subsection{Parameter Selection}

The fragment length $d$ is critical. Phase 1 requires a fragment covering at least one layer-0 block to estimate position. In Phase 2, we prefer fragments that complete blocks to verify hashes early. Incomplete blocks delay verification and increase search space, causing failures. Thus, we choose the smallest feasible $d$. 

\section{Simulation Results}
\label{sec:simulation}

We evaluate the proposed coding schemes with breakage parameters $\alpha \in \{0.05, 0.07, 0.10\}$ and substitution probabilities $p_s \in \{0.4\%, 0.9\%, 1.8\%, 5\%\}$. Efficiency is maintained by using IEEE 802.16e WiMAX and IEEE 802.22 WRAN LDPC codes \cite{rptu_wimax_ldpc} with adaptive rates and lengths (Table~\ref{tab:ldpc_rates}). Decoding is performed via the product-sum algorithm (50 iterations) for index-free codes \cite{Roffe_LDPC} and the min-sum algorithm (100 iterations) for index-based codes \cite{AFF3CT}.

\begin{table}[h]
    \centering
    \begin{tabular}{c|c|c|c|c}
         $p_s$ & $0.4\%$ & $0.9\%$ & $1.8\%$ & $5\%$ \\
        \hline
        LDPC Rate & $0.83$    & $0.75$  & $0.66$  & $0.5$  \\
    \end{tabular}
    \begin{tabular}{c|c|c|c}
         $\alpha$ & $0.05$ & $0.07$ & $0.10$ \\
        \hline
        LDPC Length & $1152$    & $576$  & $384$  \\
    \end{tabular}
    \caption{LDPC code rates and lengths for different substitution probabilities and breakage parameters.}
    \label{tab:ldpc_rates}
    \vspace{-5ex}
\end{table}

Reconstruction relies on a beam search where the number of beams and maximum number of iterations on the beams are adapted to the channel parameters as can be seen in Tables II and III in the appendix of the extended version on arXiv.
Failures were solely due to search space exhaustion (limited number of beams or search limits on the beams). Namely, while constructing partial assemblies, many wrong ones have low violations count, which results in the correct partial assembly being flushed out of the beam. In this case we do not retrieve the correct assembly, and these are the only failure cases we observed. 
The overall beam search complexity is $O\!\left(I_{\max} \cdot |\mathcal{F}| \cdot \bigl(|\mathcal{F}| \cdot n + |\mathcal{S}| \cdot (n + B_{\max})\bigr)\right)$, where $I_{\max}$ is the iteration limit.
Importantly, no incorrect codewords were returned.

\subsection{Marker Scheme Comparison}

\begin{figure}[h]
\centering
\begin{tikzpicture}
\begin{axis}[
    ybar,
    bar width=0.3cm,
    width=\columnwidth,
    height=3.5cm,
    ylabel={Error Rate},
    symbolic x coords={Marker, Stride 1, Stride 2, Block},
    xtick=data,
    ymin=0,
    every node near coord/.append style={font=\footnotesize, rotate=90, anchor=west},
    nodes near coords,
    enlarge x limits=0.2,
    enlarge y limits={upper, value=0.4},
    axis x line*=bottom,
    axis y line*=left,
    legend style={at={(0.5,-0.35)}, anchor=north, legend columns=-1, font=\footnotesize},
]
\addplot coordinates {(Marker, 0.0018) (Stride 1, 0.0024) (Stride 2, 0.0014) (Block, 0.0017)};
\addplot coordinates {(Marker, 0.0027) (Stride 1, 0.0069) (Stride 2, 0.0074) (Block, 0.0055)};
\legend{$p_s = 0.009$, $p_s = 0.018$}
\end{axis}
\end{tikzpicture}
\caption{Comparison of frame error rates (FER) across different marker schemes. \\[-6ex]}
\label{fig:marker_comparison}
\end{figure}
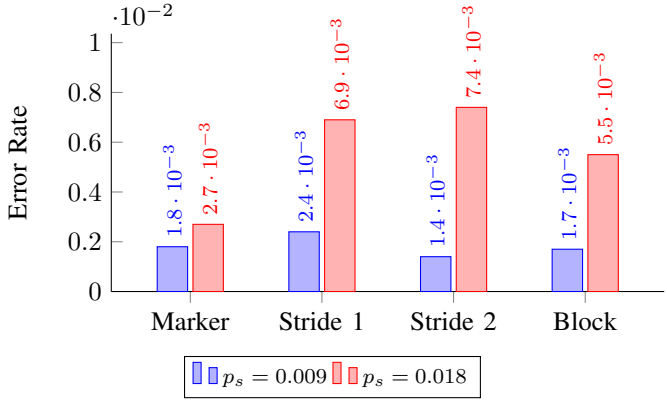

We first compare the four marker strategies (Section~\ref{sec:nested_hash}) using $n=1264$ and $\alpha=0.05$. As shown in Fig.~\ref{fig:marker_comparison}, a trade-off exists: static markers are more robust at higher substitution probabilities ($p_s=1.8\%$), whereas the data-dependent LSH scheme (specifically Stride 2) prevails at lower noise levels ($p_s=0.9\%$). This suggests that LSH offers better efficiency when data integrity is higher, while static markers provide necessary anchors in noisier conditions.
\vspace{-3ex}

\subsection{Achievable Rates}

\begin{figure}[h]
\centering
\begin{tikzpicture}
\begin{axis}[
    width=\columnwidth,
    height=5cm,
    xlabel={$\alpha$},
    xlabel style={anchor=west},
    ylabel={Code Rate},
    xmin=0.04, xmax=0.11,
    ymin=0.2, ymax=0.95,
    xtick={0.05, 0.07, 0.1},
    xticklabel style={/pgf/number format/.cd, fixed, precision=2},
    legend pos=south west,
    legend style={font=\footnotesize},
    grid=major,
    axis x line*=bottom,
    axis y line*=left,
]
\addplot[color=blue, dashed, thick, mark=none] coordinates {
    (0.05, 0.92) (0.07, 0.90) (0.1, 0.88)
};
\addlegendentry{$C(p_s = 0.4\%)$}

\addplot[color=blue, mark=*, thick] coordinates {
    (0.05, 0.78) (0.07, 0.71) (0.1, 0.57)
};
\addlegendentry{$p_s = 0.4\%$}

\addplot[color=red, dotted, thick, mark=none] coordinates {
    (0.05, 0.88) (0.07, 0.86) (0.1, 0.84)
};
\addlegendentry{$C(p_s = 0.9\%)$}

\addplot[color=red, mark=square*, thick] coordinates {
    (0.05, 0.68) (0.07, 0.63) (0.1, 0.50)
};
\addlegendentry{$p_s = 0.9\%$}

\end{axis}
\end{tikzpicture}
\\[-2ex]
\caption{Code rate vs.\ $\alpha$ for different substitution error probabilities in marker codes without indexing. \\[-6ex]}
\label{fig:rate_vs_alpha_index_marker}
\end{figure}
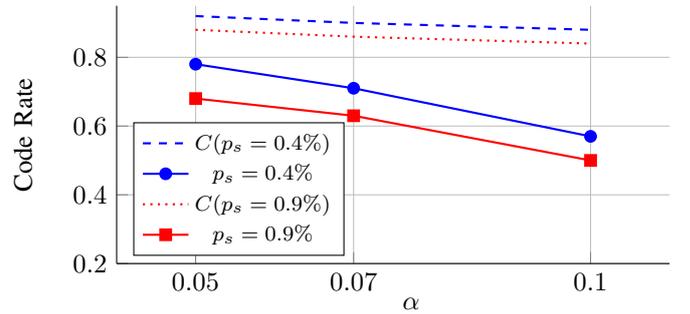

Our schemes successfully reconstructed codewords with probability $>99\%$ in all tested scenarios. Figures~\ref{fig:rate_vs_alpha_hash} and~\ref{fig:rate_vs_alpha_index_marker} display the achievable rates against theoretical capacity \cite{Lit8}. Relaxing search constraints consistently allows the algorithm to recover failed codewords, indicating that the limitation in the recovery rates lies in search complexity rather than code construction. 

We tested both coding schemes for $\alpha = 0.05$ and $p_s \in \left\{ 0.004, 0.009, 0.018, 0.05 \right\}$, and for these settings we sought sets of the schemes' parameters 
that maximize the code rate while still ensuring reconstruction probability $> 99\%$ and acceptable decoding run-time complexity.
The obtained code rates are depicted in Fig.~\ref{fig:rate_vs_alpha_hash}.

The nested coding scheme without indices was also tested for $\alpha \in \{ 0.05, 0.07, 0.1 \}$ and $p_s \in \{0.004, 0.009 \}$. The code rates achieving a reconstruction probability of $ > 99\%$ are depicted in Fig.~\ref{fig:rate_vs_alpha_index_marker}.

The detailed parameter settings for all simulations conducted can be found in Tables II and III in the appendix of the extended version on Arxiv. 

\begin{figure}[h]
\centering
\begin{tikzpicture}
\begin{axis}[
    width=\columnwidth,
    height=4.5cm,
    xlabel={$p_s$},
    ylabel={Code Rate},
    xlabel style={anchor=west, xshift=-0.4cm},
    xmode=log,
    xmin=0.003, xmax=0.06,
    ymin=0.2, ymax=0.95,
    xtick={0.004, 0.009, 0.018, 0.05},
    xticklabels={0.4\%, 0.9\%, 1.8\%, 5\%},
    scaled x ticks=false,
    legend pos=south west,
    legend style={font=\footnotesize},
    grid=major,
    axis x line*=bottom,
    axis y line*=left,
]
\addplot[color=black, dashed, thick, mark=none] coordinates {
    (0.004, 0.91) (0.009, 0.88) (0.018, 0.82) (0.05, 0.66)
};
\addlegendentry{Capacity}

\addplot[color=blue, mark=*, thick] coordinates {
    (0.004, 0.74592) (0.009, 0.66666) (0.018, 0.57657) (0.05, 0.4)
};
\addlegendentry{with indexing}

\addplot[color=red, mark=square*, thick] coordinates {
    (0.004, 0.78) (0.009, 0.70) (0.018, 0.60) (0.05, 0.44)
};
\addlegendentry{without indexing}

\end{axis}
\end{tikzpicture}
\\[-2ex]
\caption{Code rate vs.\ $p_s$ for $\alpha = 0.05$ in marker codes with and without indexing. \\[-5ex]}
\label{fig:rate_vs_alpha_hash}
\end{figure}
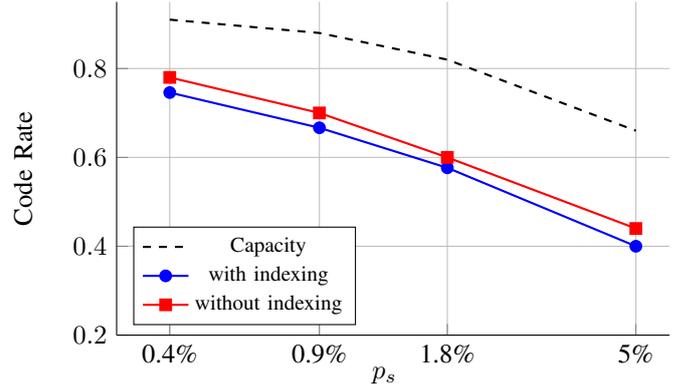

\section{Conclusion}

In this work we presented coding schemes for the noisy torn paper channel. Our analysis indicates that in noisy environments, reassembly becomes primarily a computational challenge. Notably, our decoder produced no incorrect codewords; failures resulted solely from timeouts or search space limits. In particular, the LSH scheme was demonstrated to be robust to errors for it preserves fragments connectivity even in the presence of minor corruption, a case in which high-avalanche hash functions would fail.

Our analysis focused on the low error rate regime which is guaranteed when using sophisticated DNA synthesis and sequencing technologies. We lastly leave the following tasks for future work: (1) Delving more into theoretical analysis by estimating the failure rate as a function of $B_{max}, p_s$, and $\alpha$. (2) Improving the coding scheme to tackle the rate-capacity gap depicted in Figure~\ref{fig:rate_vs_alpha_index_marker}. We believe this gap results from the redundant use of markers or the weaknesses of the hash functions. (3) Extending the evaluation to arbitrary breakage distributions, as studied by Ravi et al.~\cite{Lit8}. (4) Extending the error model to account for insertions and deletions, which would bring our schemes closer to practical DNA storage conditions. (5) Testing these algorithms on experimental datasets, like~\cite{GrassChallenges} to investigate validating practical applicability.

% Bibliography

\end{document}